\documentstyle[twocolumn,prl,aps,epsf,rotate]{revtex}
\begin{document}
\twocolumn[
\hsize\textwidth\columnwidth\hsize\csname@twocolumnfalse\endcsname

\preprint{Submitted to \prl}
\title{Flux Line Channels and Reduced Hall Effect in Impure 2D Superconductors}
\author{Kenneth Holmlund}
\address{Department of Theoretical Physics\\
Ume{\aa} University\\
S-901 87 Ume\aa, Sweden\\
Kenneth.Holmlund@TP.UmU.SE}
\date{\today ; Submitted to \prl}
\maketitle
\begin{abstract}
We have studied the dynamics of flux-lines in an
impure two-dimensional superconductor using numerical Brownian
dynamics.
In the dirty pinning limit and for small center-of-mass
velocities the flux lines prefer traveling in narrow
channels uniquely determined by the pinning configuration.
At a certain critical drag force there is a dynamic transition from
this glassy state into a more ordered crystalline state, in which
the channel structure vanishes.
The channels are studied by monitoring the particle trajectories
and the spatial distribution of vortices.
By explicitly including a Magnus force in the Langevin equations of
motion we induce a Hall field and we find that the effective Hall angle
is strongly reduced by the formation of channels, i.e. the channels
introduce a transverse energy barrier that guides the motion, predominantly
in a direction parallel to the external force.
The transverse depinning
force is larger than the depinning force in the direction
of the drag and thus the Hall field actually vanishes at finite velocities.
We propose that the Hall angle can serve as an order parameter
for experimental detection of the glass to crystal transition.
Our results are in excellent agreement
with recent predictions made by Giamarchi and Le Doussal.\cite{GiLeD95}
\end{abstract}
]
The flux line lattice (FLL) of thin film and high-T$_c$ superconductors has
been
found to exhibit many interesting properties of which the dynamic phase diagram
involving both plastic, crystalline, liquid and glassy states is
one.\cite{KoVi94}
In fact this complexity is to be expected since there in the system are four
strongly
competing energy scales; thermal fluctuations, vortex-vortex
interaction, vortex-pinning interaction and the external field. In addition
the impurities can be realized in many different ways by varying the density,
distribution, range and strength of the interaction potential, in order to
model both natural and artificially created impurities.

Much effort has been put into characterizing the nature of the dynamic
phase diagram and the dynamic phase transitions.\cite{Bra95,Bla94}
This is well motivated effort since to some extent the energy dissipation
of the dynamic vortex transitions is resposible for the actual
breakdown of the superconducting state.

Koshelev and Vinokour reported computer simulations where
they monitored the hexagonal order parameter in order to characterize
the phase diagram spanned by the external drag force and the temperature
in the presence of pinning.\cite{KoVi94}
Several attempts have been made to employ perturbative methods in order
to compute the center of mass velocity as a function of the external
force, but recently Giamarchi and Le Doussal
pointed out that perturbation
theory in the inverse velocity and disorder breaks
down for these systems because some modes of the disorder are
not affected by the motion.\cite{GiLeD95}
In addition they also predicted that
the vortices should prefer traveling in elastic channels uniquely
determined by the disorder configuration.
To some extent this prediction was confirmed already before is was made by
the experiments of Tr{\"a}uble and Essmann.\cite{TrEs68}
Indications of channel-like motion  has also been seen in computer simulations
of Brass et. al. but then for a
system with an interaction potential (Gaussian) far from the
true long-range potential and at very low pinning densities.\cite{BrJeBe89}

In this letter we will use the dynamic phase diagram presented by  Koshelev
et al.\cite{KoVi94} as a starting point and inspired by the
predictions of Giamarchi et al. \cite{GiLeD95} we carry on the
analysis further.

The starting point is the over-damped (massless)
Langevin equation for a vortex $i$,\cite{BrJe89,Bra83}
\begin{eqnarray}
\label{langevin.eq}
{\bf v}_i(t) = {\bf F}_{d} & - &
\sum_{j\ne i}{\bf\nabla} U_{vv}({\bf r}_{ij}) -
\sum_k{\bf\nabla} U_{vp}({\bf r}_{ik}) \nonumber \\
 & - & a {\bf v}_i(t)\times {\hat {\bf z}} + {\bf\zeta}_i,
\end{eqnarray}
where $U_{vv}$ is the repulsive vortex-vortex
potential, $U_{vp}$ the attractive vortex-pin
potential and ${\bf\zeta}_i$ is a noise term satisfying
$\left<\zeta_{i,\alpha}(t^\prime)\zeta_{j,\alpha^\prime}(t)\right> =
2T\delta_{ij}\delta_{\alpha , \alpha^\prime}\delta(t-t^\prime)$
where $\alpha$,$\alpha^\prime$ refer to Cartesian components and $T$ is the
temperature.  ${\bf F}_{d}$ is an external drag
force which corresponds to the applied current in the
experimental situation and $a$ is a coefficient that governs the
strength of the Magnus force (i.e. the fourth term on the l.h.s.
is the Magnus force).

The equations of motion are simulated using the method originally
described by Ermak\cite{Erm75} and later used for vortices
by  Brass et. al.\cite{BrJe89}, using a numerical time step,
$\Delta t = 0.1$.

The vortex system is simulated at intermediate magnetic fields,
for which the two-dimensional vortex interaction potential is
purely logarithmic,
$U_{vv}(r) = \ln {r/a_0}$,\cite{Vi69}
where $a_0$ is the unit measure of length here taken to be
the lattice constant of the perfect FLL. This choice of potential
also defines the unit of energy. In practice we use the Ewald
summation technique, periodic boundary conditions and the minimum image
convention in order to properly handle the long range forces
as described by e.g. Frey et. al.\cite{Frey94}
The interaction potential used here is somewhat more ambitious than
the potential used by Koshelev et al.\cite{KoVi94}, which was a
force shifted logarithmic potential explicitly cut off at $3 a_0$
(i.e. not truly long-range).
Actually a potential with shorter interaction range more or less just
corresponds to a lower magnetic field in the real system.
This is due to the peculiar length scale invariant property of the
logarithm function.
With a full logarithmic interaction all $N_v(N_v-1)/2$ pair interactions
must be taken into account. Thus the present simulations are very much
more computationally demanding than would they be if using a short range
potential and neighbor lists.
In practice it however turns out that the precise choice of
interaction potential makes little difference.

The pinning is realized from a set of $N_p$ randomly
distributed attractive Gaussian potentials $U_{vp} = -A_p e^{-(r/r_p)^2}$.
The parameter set we will consider is:
$A_p = 0.006$, $r_p = 0.2$, $N_p = 9500$, $N_v=20\times24=480$,
$0.005$, $0 \le F_d \le 0.25$ and for the Magnus coefficient
we choose the value $a=0.2$, i.e. we expect the Hall angle to be
$\Phi_H \approx 11.31^o$ when the drag force is large enough for
the impurities to be neglected (or for a pure system).

\begin{figure}
\begin{picture}(0,0)%
\includegraphics{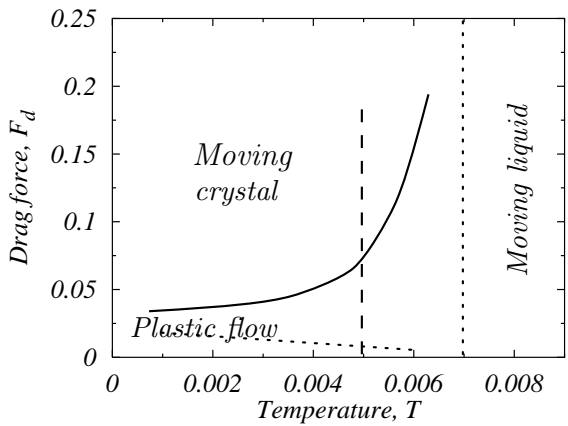}%
\end{picture}%
\setlength{\unitlength}{0.011250in}%
\begingroup\makeatletter\ifx\SetFigFont\undefined
\def\x#1#2#3#4#5#6#7\relax{\def\x{#1#2#3#4#5#6}}%
\expandafter\x\fmtname xxxxxx\relax \def\y{splain}%
\ifx\x\y   
\gdef\SetFigFont#1#2#3{%
  \ifnum #1<17\tiny\else \ifnum #1<20\small\else
  \ifnum #1<24\normalsize\else \ifnum #1<29\large\else
  \ifnum #1<34\Large\else \ifnum #1<41\LARGE\else
     \huge\fi\fi\fi\fi\fi\fi
  \csname #3\endcsname}%
\else
\gdef\SetFigFont#1#2#3{\begingroup
  \count@#1\relax \ifnum 25<\count@\count@25\fi
  \def\x{\endgroup\@setsize\SetFigFont{#2pt}}%
  \expandafter\x
    \csname \romannumeral\the\count@ pt\expandafter\endcsname
    \csname @\romannumeral\the\count@ pt\endcsname
  \csname #3\endcsname}%
\fi
\fi\endgroup
\begin{picture}(295,206)(33,586)
\put(108,632){\makebox(0,0)[lb]{\smash{\SetFigFont{11}{13.2}{it}Plastic flow}}}
\put(137,712){\makebox(0,0)[lb]{\smash{\SetFigFont{12}{14.4}{it}Moving}}}
\put(138,696){\makebox(0,0)[lb]{\smash{\SetFigFont{11}{13.2}{it}crystal}}}
\put(293,660){\makebox(0,0)[lb]{\smash{
\SetFigFont{11}{13.2}{it}Moving liquid
}}}
\end{picture}
\caption{
\label{fig.phase}
The dynamic phase diagram of the vortex system.
The dashed line starting at $T=0.005$ and $F_d=0$ is
studied in the simulations. The vertical dotted line at $T=0.007$ is
the absolute melting transition.\protect\cite{Cai82} The lower horizontal
(dotted) curve displays the plastic-elastic crossover
and the upper curve (solid) shows the transition from from glass-like
state into the ordered crystal state.
}
\end{figure}

The schematic phase diagram sketched in Fig. \ref{fig.phase} is by now well
established.\cite{KoVi94}

By varying the drag force $F_d$, in our simulation we will follow the
vertical line starting at $F_d = 0$ and $T=0.005$.

\begin{figure}
\begin{picture}(0,160)(60,60)
\includegraphics{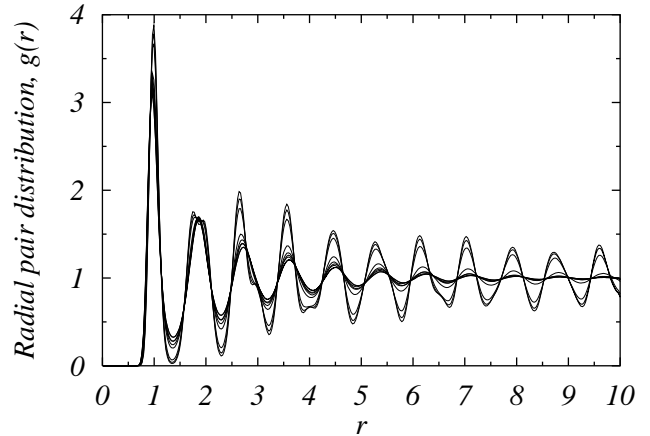}
\end{picture}\\
\caption{
\label{fig.pairc}
The dynamic glass to crystal transition is clearly and intuitively
visualized by the vortex-vortex pair distribution functions.
The set of curves with smaller amplitude correspond to the drag
forces $F_d = 0.03, 0.04, 0.05, 0.06$, below the transition,
and the curves with larger amplitudes are for $F_d=0.07, 0.08, 0.1$
above the transition, resplectively. The transition occurs at,
$F_d \approx 0.062$.
}
\end{figure}

\begin{figure}

\begin{picture}(100,139)(-45,0)
\includegraphics{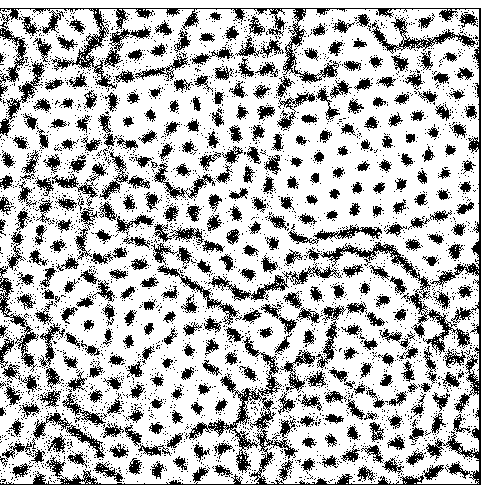}
\end{picture}\\
\begin{picture}(100,139)(-55,0)
\includegraphics{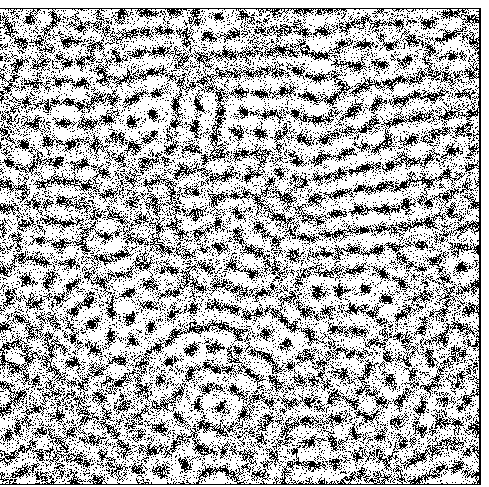}
\end{picture}\\
\begin{picture}(100,139)(-55,0)
\includegraphics{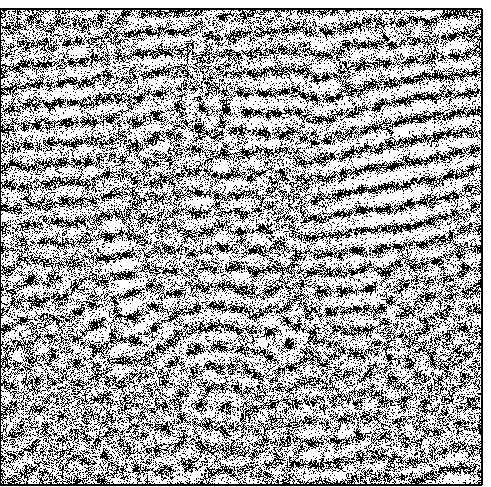}
\end{picture}\\
\begin{picture}(100,139)(-55,0)
\includegraphics{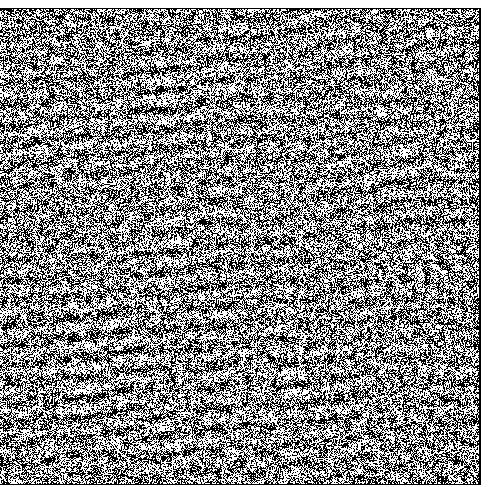}
\end{picture}\\
\caption{
\label{fig.traj}
Time averaged density maps show the preferred spots for a
vortex and illustrates how the vortices prefer traveling
in elastic channels. Figures a) - d) (from top to bottom)
correspond to different drag forces,
$F_d = 0.002, 0.015, 0.050, 0.080$.
}
\end{figure}

As also was noted by Koshelev et al.\cite{KoVi94} the transition
from the glassy state into the more ordered crystalline phase
is sharp. The transition can actually be seen directly in the
vortex pair distribution function which is plotted in Fig. \ref{fig.pairc}.
The set of curves with smaller amplitudes correspond to drag forces
below the transition whereas the set of curves with large amplitudes
correspond to forces above the transition. Using this somewhat crude
order parameter we determine the transition point to be located
at $F_d \approx 0.062$ for this particular pinning configuration.
The formation of channels is clearly illustrated in Figs.\ref{fig.traj}~a-d
where every 175th configuration out of totally 100000 simulation steps
is plotted. This gives a time averaged map of the spatial
distribution of vortices in the simulation cell. Each dot
represents a vortex and thus the darkest areas are those most favored
by the the flowing vortices.
The same pinning configuration but different starting conditions were
used to produce these plots.

\begin{figure}[ht]
\begin{picture}(100,175)(73,60)
\includegraphics{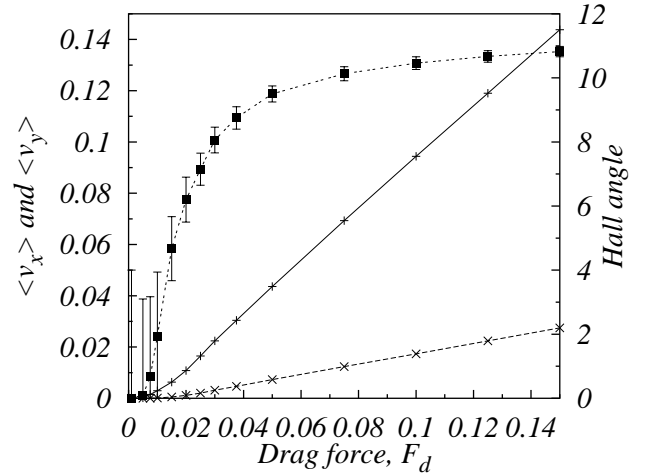}
\end{picture}
\\
\caption{
\label{fig.vvsFd}
The average center of mass vortex velocities in the transverse (broken)
and horizontal (solid) directions are plotted as a function of the
drag force $F_d$. The corresponding Hall angle shows that the transverse
flow is strongly reduced at low velocities due to the presence of the
elastic channels.
}
\end{figure}

Starting at small forces, $F_d = 0.002$ (Fig.\ref{fig.traj}a), we find that 
most vortices are pinned to the background and the total flow is almost
completely suppressed by the pinning (see also Fig.\ref{fig.vvsFd}).
The few vortices that actually do move, creep in channels
between the pinned chunks of the lattice.
For $F_d = 0.015$ (Fig.\ref{fig.traj}b). the channel structure is
clear. Very few vortices are pinned for long times but
the mobility is still strongly reduced by the pinning.
More close to the transition, for $F_d = 0.05$, the channel
structure is still present but the vortices have a tendency
to "tunnel" between the channels. Finally for $F_d = 0.08$,
above the transition, only a very weak channel structure can
be discerned.

It is clear that the channel structure is indeed unique for
each pinning configuration since runs with very
different initial conditions eventually end up with the
same flow patterns for a given configuration of pins.

The dependence of the transverse and perpendicular center-of-mass
velocities with external force (i.e. the IV-characteristics)
is plotted in Fig. \ref{fig.vvsFd}. In the same diagram
also the effective Hall angle is plotted and it is clear
that the Hall field becomes strongly reduced as soon as the channels
start to form. Interestingly enough the angle
is slightly reduced even at velocities far into the "Moving crystal"
region of Fig.\ref{fig.phase}.
These velocity versus force curves have been averaged over 15
or more realizations of the disorder.

The strong dependence of the Hall angle with the drag force should
be encouraging for designing experiments since it may serve as an order
parameter for the glass to crystal transition. It would also be
interesting to make more quantitative comparisons with the real
system by applying the emperical Corter-Casimir formulas to the
model used here, in order to obtain the correct temperature and
magnetic field dependencies.\cite{deG66,BrJe89}
It has been argued that the vortex lattice tries
to minimize it's power dissipation by reorienting in the
direction of the drag force.\cite{SmiHa73}
This has not yet been verified neither experimentally nor
in simulations although the effect probably is of great
importance.

The glass to crystal transition is very sharp in the
structural and geometrical order parameters and should therefore
be accompanied by a strong buildup of washboard modes, i.e.
harmonic components in the center-of-mass velocity.
We have preliminary simulational data that verify that not only
the fundamental mode but also higher order frequencies are
important in the spectrum of the center-of-mass velocity.\cite{KH96}
Coupling of these modes may enhance the strength of the
transition and drive the system into a more ordered state.
This and many other issues will be dealt with in a forthcoming
publication.\cite{KH96}

In conclusion we have shown that the flux-lines of a 2D superconductor
at low velocities flow in elastic channels that strongly suppresses the
Hall field and that the Hall angle vanishes already at finite velocities.
The flow patterns are uniquely determined by the pinning configuration.

The author gratefully acknowledges Prof. Petter Minnhagen, Dr. Mats Nyle\'n,
Dr. Peter Olsson and Dr. Andrei Shelankov for fruitful discussions.
Simulations have been performed on DEC alpha and HP workstations.


\begin{thebibliography}{99}
\bibitem{GiLeD95}
T. Giamarchi and P. Le Doussal,
cond-mat/9512006.
\bibitem{KoVi94}
A. E. Koshelev and V. M. Vinokur,
Phys. Rev. Lett. {\bf 73}, 3580 (1994).
\bibitem{Bra95}
For a review see e.g. E. H. Brandt, supr-con/9506003.
\bibitem{Bla94}
For a review see e.g.
G. Blatter et. al., Rev. Mod. Phys. {\bf 66}, 1125 (1004).
\bibitem{TrEs68}
H. Tr{\"a}uble and U. Essmann, Phys. Stat. Sol. {\bf 25}, 395 (1968).
\bibitem{BrJeBe89}
A. Brass, H.J. Jensen, and A. J. Berlinsky, Phys. Rev. B {\bf 39}, 102 (1989).
\bibitem{BrJe89}
A. Brass and H. J. Jensen, Phys. Rev. B {\bf 39}, 9587 (1989)
\bibitem{Bra83}
E. H. Brandt, J. Low. Temp. Phys. {\bf 53}, 41 (1983).
\bibitem{Erm75}
D. L. Ermak, Jou. Chem. Phys. {\bf  62}, 4189 (1975).
\bibitem{Frey94}
E. Frey, D. R. Nelson, and
D. S. Fisher, Phys. Rev. B {\bf 49}, 9723 (1994).
\bibitem{Vi69}
W. F. Vinen, in {\it Superconducting}, edited by R.D. Parks (Marcel
Dekker, New York, 1969), Vol. 2.
\bibitem{Cai82}
J. M. Caillol et al., J. Stat. Phys. {\bf 28}, 325 (1982).
\bibitem{deG66}
See for example: P. G. de Gennes, {\it Superconductivity of Metals
and Alloys}, Benjamin New York, (1966).
\bibitem{SmiHa73}
A. Schmid and W. Hauger, J. Low Temp. Phys. {\bf 11} 667 (1973).
\bibitem{KH96}
K. Holmlund, in preparation.
\end{thebibliography}
\end{document}